\documentclass[conference,compsoc]{IEEEtran}

\usepackage{url,hyperref,lineno,microtype,subcaption}
\usepackage[onehalfspacing]{setspace}
\usepackage{tikz}

\newcommand*\circled[1]{\tikz[baseline=(char.base)]{
    \node[shape=circle,draw,inner sep=2pt] (char) {#1};}}

\def\addvalue#1#2{\expandafter\gdef\csname note@colors@#1\endcsname{#2}}
\def\usevalue#1{\csname note@colors@#1\endcsname}
\def\setnotecolor#1#2{\addvalue{color-#1}{#2}}

\newcommand\draft{}
\setnotecolor{AM}{Green}
\setnotecolor{MH}{Violet}
\setnotecolor{Kyle}{Red}
\setnotecolor{Outline}{RubineRed}
\setnotecolor{mturilli}{Orange}
\setnotecolor{Ian}{blue}
\setnotecolor{AW}{SpringGreen}
\setnotecolor{Aymen}{Purple}
\setnotecolor{Mikhail}{Red}

\newcommand{\note}[2]{\par{\color{\usevalue{color-#1}} #1: #2}}
\newcommand{\inlinenote}[2]{~{\color{\usevalue{color-#1}} #1: #2}}
\newcommand{\resolvednote}[2]{}
\newcommand{\resolvedinlinenote}[2]{}

\unless\ifdefined\draft
    \renewcommand{\note}[2]{}
    \renewcommand{\inlinenote}[2]{}
\fi

\begin{document}

\title{ExaWorks Software Development Kit: A Robust and Scalable Collection of Interoperable Workflows Technologies}



\author{
    \IEEEauthorblockN{
        Matteo Turilli\IEEEauthorrefmark{1}\IEEEauthorrefmark{2},
        Mihael Hategan-Marandiuc\IEEEauthorrefmark{3}\IEEEauthorrefmark{4},
        Mikhail Titov\IEEEauthorrefmark{1},
        Ketan Maheshwari\IEEEauthorrefmark{5},
        Aymen Alsaadi\IEEEauthorrefmark{2}\\
        Andre Merzky\IEEEauthorrefmark{6},
        Ramon Arambula\IEEEauthorrefmark{7},
        Mikhail Zakharchanka\IEEEauthorrefmark{7},
        Matt Cowan\IEEEauthorrefmark{1},
        Justin M. Wozniak\IEEEauthorrefmark{4}\\
        Andreas Wilke\IEEEauthorrefmark{3}\IEEEauthorrefmark{4},
        Ozgur Ozan Kilic\IEEEauthorrefmark{1},
        Kyle Chard\IEEEauthorrefmark{3}\IEEEauthorrefmark{4},
        Rafael Ferreira da Silva\IEEEauthorrefmark{5},
        Shantenu Jha\IEEEauthorrefmark{1}\IEEEauthorrefmark{2} and
        Daniel Laney\IEEEauthorrefmark{7}
    }
    \IEEEauthorblockA{
        \IEEEauthorrefmark{1}Brookhaven National Laboratory, Upton, NY, USA
    }
    \IEEEauthorblockA{
        \IEEEauthorrefmark{2}Rutgers University, New Brunswick, NJ, USA
    }
    \IEEEauthorblockA{
        \IEEEauthorrefmark{3}University of Chicago, Chicago, IL, USA
    }
    \IEEEauthorblockA{
        \IEEEauthorrefmark{4}Argonne National Laboratory, Lemont, IL, USA
    }
    \IEEEauthorblockA{
        \IEEEauthorrefmark{5}Oak Ridge National Laboratory, Oak Ridge, TN, USA
    }
    \IEEEauthorblockA{
        \IEEEauthorrefmark{6}Incomputable LLC, Highland Park, NJ, USA
    }
    \IEEEauthorblockA{
        \IEEEauthorrefmark{7}Lawrence Livermore National Laboratory, Livermore, CA, USA
    }
}

\maketitle

\begin{abstract}
Scientific discovery increasingly requires executing heterogeneous scientific workflows on high-performance computing (HPC) platforms. Heterogeneous workflows contain different types of tasks (e.g., simulation, analysis, and learning) that need to be mapped, scheduled, and launched on different computing. That requires a software stack that enables users to code their workflows and automate resource management and workflow execution. Currently, there are many workflow technologies with diverse levels of robustness and capabilities, and users face difficult choices of software that can effectively and efficiently support their use cases on HPC machines, especially when considering the latest exascale platforms. We contributed to addressing this issue by developing the ExaWorks Software Development Kit (SDK). The SDK is a curated collection of workflow technologies engineered following current best practices and specifically designed to work on HPC platforms. We present our experience with (1) curating those technologies, (2) integrating them to provide users with new capabilities, (3) developing a continuous integration platform to test the SDK on DOE HPC platforms, (4) designing a dashboard to publish the results of those tests, and (5) devising an innovative documentation platform to help users to use those technologies. Our experience details the requirements and the best practices needed to curate workflow technologies, and it also serves as a blueprint for the capabilities and services that DOE will have to offer to support a variety of scientific heterogeneous workflows on the newly available exascale HPC platforms.
\end{abstract}

\section{Introduction}\label{sec:intro}

Workflow systems executed at unprecedented scale are increasingly necessary to enable scientific discovery~\cite{badia2017workflows}. Contemporary workflow applications benefit from new AI/ML algorithmic approaches to traditional problems but also bring new and challenging computing requirements to the fore~\cite{da2021community}. Exascale High-performance computing (HPC) platforms can satisfy the growing need for scale but at the cost of requiring middleware with increased complexity and multiple dimensions of heterogeneity. A renewed impulse to develop workflow applications for HPC platforms meets a traditionally fragmented software ecosystem, creating several issues for the users and middleware developers.

Domain scientists who code HPC workflow applications face four main challenges: (1)~choosing a workflow system that satisfies the requirements of their use cases and their target HPC platform(s); (2)~learning to use the chosen workflow system to code and execute their applications; (3)~deploying the workflow system and its middleware stack on the target HPC platform, testing that they work as intended; (4)~maintaining their workflow application over time while ensuring that the chosen workflow system remains functional through the routine software updates of their target HPC platform.

In a fragmented ecosystem with many workflow systems and workflow-related technologies specifically designed to support scientific applications, domain scientists must review multiple software solutions with similar capabilities but diverse maturity levels, robustness, reliability, documentation, and support. Furthermore, users must learn the basic functionalities of several systems to evaluate, deploy, and test them on one or more HPC platforms. Together, that requires a significant effort involving significant human and time resources. Further, even after choosing a viable software stack, the need for portability due to the different capabilities of the HPC platforms and the availability of multiple allocations and recurrently updated machines leads to a constant process of deploying, testing and fixing both the workflow applications and the middleware that support their execution.

Given the needed resources, scientists often meet those challenges by relying on `word of mouth,' repeating the choices made by colleagues even if sub-optimal. Alternatively, users resort to coding new single-point solutions, furthering the balkanization of the existing software ecosystem. Ultimately, both options make it difficult, if not impossible, to express the potential of HPC scientific workflow applications to support innovation and discovery. We propose that overcoming that limitation requires a novel approach to make available a selected set of tools that can provably support the execution of scientific workflow applications on the largest HPC platforms currently available to the scientific community. Further, we need a way to grow the existing software ecosystem, avoiding effort duplication while improving scalability, portability, usability, and reliability and lowering the learning curve and the maintenance effort.

This paper describes the lessons learned while delivering ExaWorks~\cite{al2021exaworks}, a project designed to implement the approach described above. Specifically, this paper focuses on the ExaWorks Software Development Kit (SDK), an effort to select a complementary set of workflow technologies, make their integration possible while maintaining their individual capabilities, testing them via continuous integration (CI), making the result of those tests publicly available, and documenting them in a way that helps to lower the access barriers when used on HPC platforms. This work serves not only as a blueprint for realizing the full potential of modern scientific workflow applications but also as a critical overview of the pros and cons of that approach, as experienced during a three-year-long project.

The rest of the paper is organized as follows. In \S\ref{sec:related}, we briefly review similar efforts, outlining similarities and relevant differences. \S\ref{sec:components} describes each software tool currently included in the ExaWorks SDK. In \S\ref{sec:integration}, we detail one of the main technical achievements of the ExaWorks project: integrating the SDK software components to provide users with a variety of new capabilities while reducing to a minimum the new code. \S\ref{sec:sucess} offers an overview of exemplar success stories about using SDK in real-world scientific research. \S\ref{sec:testing} describes the other two main deliverables of the ExaWork project: a testing infrastructure based on a CI infrastructure for Department of Energy (DOE) HPC platforms and a monitoring infrastructure based on a public dashboard, collecting the results of the CI runs. In \S\ref{sec:docs}, we illustrate the system we built to document the SDK based on containerized executable tutorials. Finally, \S\ref{sec:conclusions} summarizes SDK's characteristics, the lessons we learned while designing and developing ExaWorks SDK, and the challenges that still lie ahead.

\section{Related Work}\label{sec:related}

In this section, we briefly discuss some of the projects undertaken in the community and how the SDK follows in the steps of those experiences. The Workflows Community Initiative (WCI)~\cite{wci-web} represents a pioneering collaboration to propel advancements in workflow systems and associated technologies. This initiative unites diverse stakeholders, including researchers, developers, and industry practitioners, to enhance dialogue and cooperation on pivotal aspects of workflow management. These aspects encompass the design, execution, monitoring, optimization, and interoperability of workflow systems. A cornerstone of the initiative is the organization of the Workflows Community Summits~\cite{da2023workflows}. These summits are international workshops that serve as a dynamic platform for attendees to exchange insights, discuss the latest trends and challenges, and forge new partnerships. ExaWorks directly collaborated with and participated in WCI's activities, eliciting the requirements for its SDK.

SDK addresses several of the community's recommendations for the technical roadmap defined during the 2021 and 2022 summits~\cite{da2021community,da2023workflows}. SDK provides software products and technical insight for the interoperability of workflow systems, delivering integrations among all its workflow technologies (see \S\ref{sec:integration}). Indirectly, SDK software integrations also contribute to furthering the understanding of the roles played by standardization, placing the accent on integrating a diversity of programming models and interfaces instead of searching for a single encompassing solution. For example, the DOE Integration Research Infrastructure effort to standardize interoperable workflows is considered a potential venue. SDK packaging and testing infrastructure (see \S\ref{sec:testing}) contributes tangible capabilities to improve the support of workflow technologies on HPC platforms, and SDK documentation centered on containerized tutorials (see \S\ref{sec:docs}) contributes to training and education. While SDK does not contribute directly to the FAIR, AI, and Exascale challenges, its workflow technologies (see \S\ref{sec:components}) have a roadmap to develop capabilities that will contribute to addressing those challenges.

Separately to the community, ``standards'' for workflow languages have been developed, such as the Common Workflow Language~\cite{amstutz2016common} and Workflow Definition Language. Additionally, a few standard formats have been tried and tested, including JSON and YAML~\cite{blin2003reuse,deline2021glinda,ristov2021afcl}. In the previous decades, large projects such as SHIWA~\cite{korkhov2012shiwa} undertook the task of workflow interoperability by developing an interoperable representation of a workflow language that multiple workflow managers could enact. ExaWorks SDK takes a different approach in which tools with diverse user-facing application programming interfaces (API) and approaches to workflow specification are included in the SDK and, when needed, integrated with other tools that provide runtime capabilities. In that way, users can `compose' an end-to-end software stack that best fits their workflow application requirements.

Software development activities as part of the ECP project resulted in several scientific software tools that were, in turn, used by other projects that were built on top of them. For instance, AMReX~\cite{amrex-21} mesh refinement modeling forms the basis for several other tools developed on top of it, such as FerroX~\cite{ferrox-web}, ExaAM, and many more, forming a collection of tools for science. xSDK~\cite{xsdk-web} is a similar effort to ours. ExaWorks SDK is designed to take inspiration from those experiences. Still, it differs from them because integration is considered at the level of middleware components instead of focusing on the (math) libraries level. Therefore, the term `SDK' assumes a more general meaning than the one usually associated with a set of libraries that enables the writing of a new application.

Considering ECP, ExaWorks SDK is more akin to the DOE-sponsored Extreme-scale Scientific Software Stack (E4S)~\cite{heroux2023scalable}, a community effort that provides open-source software packages for developing, deploying, and running scientific applications on HPC and AI platforms. E4S provides from-source builds, containers, and preinstalled versions of a broad collection of HPC and AI software packages~\cite{e4s-web}. Compared to EC4, ExaWorks SDK takes a loosely coupled approach where tools maintain a high degree of independence and are not organized into a named distribution. For example, the SDK does not mandate a packaging format for its components, as E4S does. Further, ExaWorks SDK focuses on workflow technologies, not on those utilized by the tasks of those workflows.

Not all workflow technologies and their middleware implementations have been designed to foster integration with other tools. For example, Python packages such as Dask~\cite{dask} come prepackaged with some rudimentary API to interface with the resource manager. However, they are limited in functionality and tightly integrated with Dask. For that reason, SDK is a curated set of workflow technologies, and, as described in the next section, part of the effort goes into evaluating whether a prospective component can be integrated with other components without significant engineering effort.

\section{Core Components}\label{sec:components}

Exaworks SDK seeding technologies were chosen based on the requirements elicited by engaging with the DOE Exascale Computing Project (ECP) workflow communities. We created a survey to identify existing workflow systems efforts, both ECP-related and within the broader DOE software ecosystem. That survey helped us understand the challenges, needs, and possible collaborative opportunities between workflow systems and the ExaWorks project. We took a broad view of workflows, including automated orchestration of complex tasks on HPC systems (e.g., DAG-based and job packing), coupling simulations at different scales, adaptive/dynamic machine learning applications, and other efforts in which a variety of possibly related tasks have to be executed at scale on HPC platforms. For example, after eliciting a brief description of an exemplar workflow, we asked about internal/external workflow coordination, task homogeneity/heterogeneity, and details about the adopted workflow tools.

Alongside the outcome of more than twelve community meetings, the results of our survey highlighted the state of the art of scientific workflow in the ECP community. To summarize, many teams are creating infrastructures to couple multiple applications, manage jobs---sometimes dynamically---and
orchestrate compute/analysis tasks within a single workflow and manage data staging within and outside the HPC platforms. Overall, there is an evident duplication of effort in developing and maintaining infrastructures with similar capabilities. Further, customized workflow tools incur significant costs to port, maintain, and scale bespoke solutions that serve single-use cases on specific platforms and resources. These tools do not always interface with facilities smoothly and are complex and/or costly to port across facilities. Finally, the lack of proper software engineering methodologies leads to repeated failures, difficulty in debugging, and expensive fixes. Overall, there was agreement in the community that costs could be minimized, and quality could be improved by creating a reliable, scalable, portable software development kit (SDK) for workflows.

Based on the requirements summarized above, the ExaWorks SDK collects software components that enable the execution of scientific workflows on HPC platforms. We consider a reference stack that allows the development of scientific workflow applications, resolving the task dependencies of that application, acquiring resources for executing those tasks on a target HPC platform, and then managing the execution of the tasks on those resources. The SDK includes components that deliver a well-defined class of capabilities of that reference stack with different user-facing interfaces and runtime capabilities. Consistently, the SDK is designed to grow its components, including software systems that are open source (i.e., released under a permissive or copyleft license) and developed according to software engineering best practices, such as test-driven development, release early and often, and version control.

SDK faces a fragmented landscape of workflow technologies, with diverse systems that offer overlapping capabilities. To help reduce duplication, SDK collects systems that can be integrated but have distinguishing capabilities, expose diverse APIs, and support different programming models. By integrating these systems, SDK reduces the need to duplicate existing capabilities within each tool. Nonetheless, each tool's distinguishing capabilities can be used to support use cases that require a specific API, programming model, scaling performance, or programming language. Ultimately, SDK does not commit to a vision in which a single technology can solve all the requirements or all the use cases. Instead, we commit to fostering an ecosystem of integrable and complementary tools that can be used independently or combined, depending on the specific requirements of each use case.

\begin{figure}[ht]
     \centering
     \includegraphics[width=0.49\textwidth]{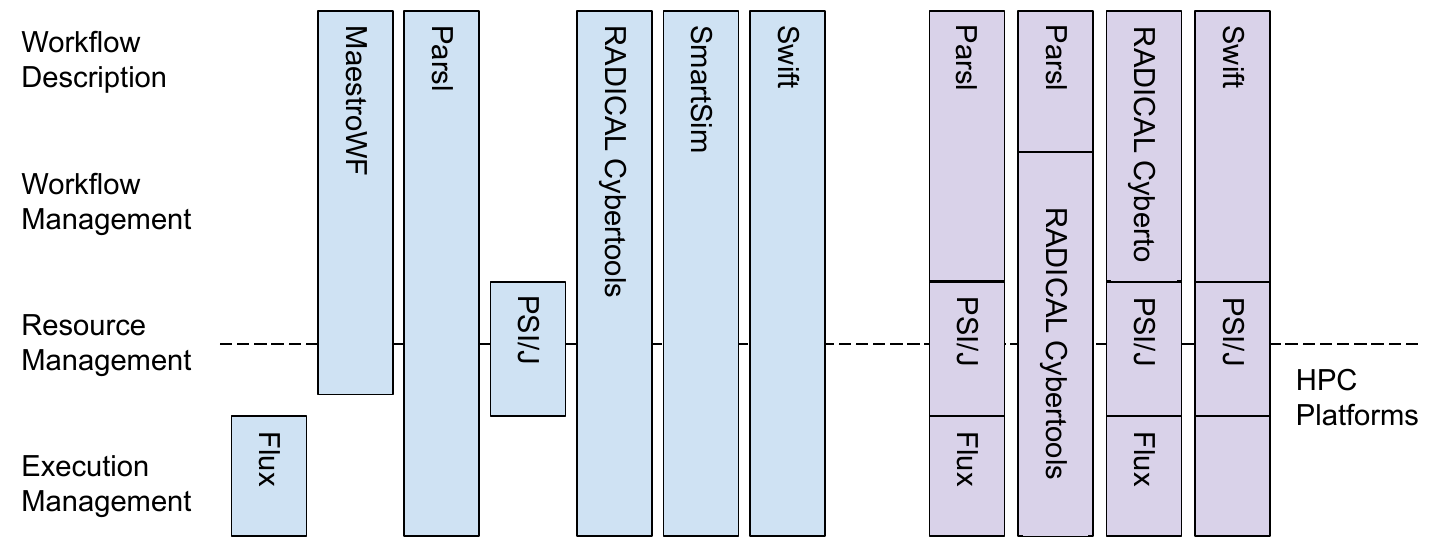}
     \caption{ExaWorks SDK reference stack (right), current components (blue boxes), and examples of integration among components (purple boxes). SDK offers a variety of interfaces, programming models, and runtime capabilities to execute scientific workflows at scale on HPC platforms.}
     \label{fig:sdk_stack}
\end{figure}

Currently, the SDK contains seven software components: Flux~\cite{flux}, MaestroWF~\cite{dinatale19maestro}, Parsl~\cite{babuji19parsl}, PSI/J~\cite{psij-23}, RADICAL Cybertools~\cite{balasubramanian2016ensemble,merzky2021design}, SmartSim~\cite{smartsim} and Swift~\cite{wozniak2013swift}. As shown in Fig~\ref{fig:sdk_stack}, each system delivers capabilities end-to-end or at a specific level of our reference stack. That way, we offer users alternative tools across or along the reference stack, depending on their specific requirements. For example, Parsl, RADICAL Cybertools, SmartSim, Swift, and, to a certain extent, MaestroWF all offer end-to-end workflow capabilities but with different programming models, application programming interfaces (API), support for HPC platforms, and scientific domain. Flux and PSI/J offer capabilities focused on resource and execution management, enabling the execution of user-defined workflows on various HPC platforms.

Further, each component is designed with sub-components, exposing well-defined interfaces. That allows us to promote integration among components, obtain new capabilities, and avoid lock-in into solutions maintained by a single team or designed to support a specific class of scientific problems and platforms. Here, we briefly describe each tool, showing how their capabilities fit the reference stack. In the next section, we detail the integration among some of these components.

\textbf{Flux} \cite{flux, AHN2020202} is a next-generation resource management and scheduling framework project under active development at LLNL. It is composed of a modular set of projects, tools, and libraries that can provide both system- and user-level resource managers and schedulers under one common software framework: system administrators and end users alike can create their instances of Flux on a set of HPC resources using the same commands and APIs to manage them according to their requirements. Furthermore, the owners of a Flux instance (e.g., a specific user who gets a set of compute nodes allocated) can spawn one or more child Flux instances that can manage a subset of the parent’s resources while specializing their services, and such a nesting can further recurse.

Flux’s fully hierarchical, customizable software framework architecture has proven effective on high-end systems, including pre-exascale systems (e.g., LLNL Sierra) as low-level service building blocks for complex workflows~\cite{dinatale19maestro}. For instance, Flux’s flexible design allows users to decide whether or not co-scheduling should be configured and lets users choose their scheduling policies (e.g., a policy optimized for high job throughput) within the scope of their instance. Further, these workflows can connect to their Flux instance and use Flux’s communication primitives such as publish-subscribe, request-reply, and push-pull, as well as asynchronous event handling to facilitate the communication and coordination between co-scheduled jobs via Flux’s well-defined, highly portable APIs.

\textbf{MaestroWF} (Maestro Workflow Conductor) allows users to define multistep workflows and automate the execution of software flows on HPC resources. It uses YAML-based study specifications to describe workflows as directed acyclic graphs that can be parameterized across multiple parameters. Maestro's study specification helps users think about complex workflows in a step-wise, intent-oriented manner that encourages modularity and tool reuse. Maestro runs in user space and does not rely on external services. Maestro is in production use at Lawrence Livermore National Laboratory by a growing user community (several dozen regular users).

\textbf{SmartSim} is a workflow library that makes it easier to use common Machine Learning (ML) libraries, like PyTorch and TensorFlow, in HPC simulations and applications. SmartSim launches ML infrastructure on HPC systems alongside user workloads and supports most HPC workload managers (e.g., Slurm, PBSPro, LSF). SmartSim also provides a set of client libraries in Python, C++, C, and Fortran. These client libraries allow users to send and receive data between applications and the machine learning infrastructure. Moreover, the client APIs enable the execution of machine learning tasks like inference and online training from within user code. The exchange of data and execution of machine learning tasks is orchestrated by a high-performance in-memory database launched and managed by SmartSim.

\textbf{Parsl} is a parallel programming library for Python that supports the definition and execution of dataflow workflows. Developers annotate Python programs with decorators, Parsl \textit{apps}, indicating opportunities for asynchronous and concurrent execution. Parsl supports two decorators: PythonApp for the execution of Python functions and BashApp to support the execution of external applications via the command line. Parsl enables workflows to be composed implicitly via data exchange between apps. Parsl supports exchanging Python objects and external files, which can be moved using various data transfer techniques. Parsl programs are portable, enabling them to be moved or scaled between resources, from laptops to clouds and supercomputers. Users specify a Python-based configuration describing how resources are provisioned and used. The Parsl runtime is responsible for processing the workflow graph and submitting tasks for execution on configured resources.

Parsl implements a three-layer architecture that makes it amenable to interoperation with other SDK components. Parsl workflows are interpreted and managed by the \textit{DataFlowKernel} (DFK). The DFK holds the dependency graph, determines when dependencies are met, and passes tasks for execution via the Executor interface. The \textit{Executor} interface, implementing Python's \texttt{concurrent.Futures} Executor, supports task-based execution, returning a Future instead of results. Parsl has several Executors designed for specific purposes, such as high throughput and extreme scale, several external executors have also been integrated as we describe in \S\ref{sec:integration}. Finally, the \textit{Provider} interface is responsible for provisioning resources from different parallel and distributed computing resources. This interface enables integration with PSI/J as a direct replacement for in-built capabilities.

\textbf{PSI/J} is an API specification for job submission management. One of its primary goals is to provide an API that abstracts access to local resource managers (LRMs), such as Slurm or PBS. PSI/J aims to replace the ad-hoc solutions found in most workflow systems and other software that require portability across multiple HPC clusters. PSI/J comes with a reference Python implementation and an infrastructure for distributed and user-directed testing. That infrastructure enables testing of PSI/J on a wide range of user-accessible platforms, centralizing and making test results publicly accessible. We used a modified version of the PSI/J testing infrastructure for the ExaWorks SDK components.

\textbf{RADICAL Cybertools (RCT)} are middleware software systems designed to develop efficient and effective tools for scientific computing. Specifically, RCT enables developing applications that can concurrently execute up to $10^5$ heterogeneous single/multi-core/GPU/node MPI/OpenMP tasks on more than twenty HPC platforms. Tasks can be implemented as stand-alone executables and/or Python functions. RCT comprises building blocks designed to work as stand-alone systems, integrated among themselves or with third-party systems. RCT enables innovative science in multiple domains, including biophysics, climate science, particle physics, and drug discovery, consuming hundreds of millions of core hours/year. Currently, SDK includes two RCT systems: RADICAL-Pilot (RP)~\cite{merzky2021design} and RADICAL-EnsembleToolkit (EnTK)~\cite{balasubramanian2016ensemble}. Both implemented as independent Python modules, RP is a pilot enabled~\cite{turilli2018comprehensive,luckow2012p} runtime system, while EnTK is a workflow engine designed to support the programming and execution workflows with ensembles of heterogeneous tasks.

\textbf{Swift/T} is a workflow system for single-site workflows. It is based on an automatically parallelizing programming language and an MPI-based runtime. The goal of Swift/T is to efficiently manage workloads consisting of many compute-intensive tasks, such as scientific simulations or machine learning training runs, and distribute them at a fine-grained level across the CPUs or GPUs of the site. The language aspect of Swift/T allows the user to define executions in terms of Python, R, or shell script fragments and then set up a data dependency structure that specifies the order in which data is created. Tasks are executed, possibly including loops, recursive function calls, and other complex patterns. The language has a familiar C or Java-like syntax but automatically provides concurrency within a dataflow control paradigm. Thus, loops and many other syntax features are automatically parallelized. The developer can use the language to launch tasks, manage workflow-level data, and even launch subordinate MPI jobs using multiple mechanisms. Developers can leverage standard mechanisms to use GPUs, node-local storage, and other advanced features.

\section{Component Integration}\label{sec:integration}

As seen in \S\ref{sec:intro}, the landscape of the software that supports the execution of scientific workflows is fragmented into tools with similar capabilities, often without proper maintenance and support. The SDK contributes to reducing that fragmentation by maintaining a collection of adequately engineered, production-grade workflow tools with a proven track record and promoting integration among those tools. The underlying idea is to avoid reimplementing already available capabilities, instead spending the resources on designing and coding integration layers between the technologies with diverse capabilities. While the idea is simple, its actual implementation is challenging. Software systems designed by independent engineering teams are not thought to be compatible at the abstraction and implementation levels.

Our integration experience showed a discrete homogeneity of abstractions among software designed to support scientific workflows. For example, most tools share analogous Task abstractions, assumptions about data dependencies among tasks, and, to some extent, the internal representation of computing resources and how they relate to computing tasks. Most differences were found at the interface and runtime level, where each tool implements distinctive designs and capabilities. SDK tools adopt similar best engineering practices (as most software is designed for production these days), contributing to a certain degree of homogeneity. For example, most SDK tools adopt designs based on well-defined APIs, connectors, and adaptors. Finally, implementation-wise, some tools adopted similar technologies, such as Python and ZeroMQ.

On those bases, integration points were clearly to be found at the level of the connectors/adapters and primarily based on translating internal representations of tasks, their data, and their resource requirements. Overall, we could integrate user-facing capabilities with various runtime capabilities, requiring minimal code to be written, mainly to implement translation layers among well-defined interfaces or connectors from existing base classes. Integrating models of resource representations was more challenging, especially concerning resource acquisition capabilities. Sometimes, that required extending existing user-facing interfaces to allow a unified definition of resource requirements, and other times, modifying the data structures and methods used to manage resources at runtime. Even accounting for such code extensions and modifications, integration proved straightforward from both a design and implementation perspective.

Overall, the experience gained with the development of SDK shows that to mitigate and reduce fragmentation of the landscape of scientific workflow technologies, tools need to be designed following engineering best practices and established patterns for distributed systems, adopt common abstractions, and utilize common patterns and separation of concerns for the communication and coordination protocols. Perhaps unsurprisingly, by doing all the above, tools are more likely to be able to be integrated with future third-party tools. In the remaining section, we expand this summary by describing the integration between Parsl and RADICAL-Pilot, Parsl and PSI/J, RADICAL-Pilot and Flux, RADICAL-Pilot and PSI/J, and Swift/T and PSI/J. Together, those integrations detail the lesson we learned by delivering SDK.

\subsection{RADICAL-Pilot and Parsl Integration}

Integrating RADICAL-Pilot (RP) and Parsl allows two tools developed by different research groups to work seamlessly together, delivering new capabilities that were not available without their integration. RP and Parsl integration is based on a loosely coupled design, in which RP becomes an executor of Parsl (called RPEX), and both systems send, receive, and share tasks~\cite{alsaadiParslRP2022}. RP offers capabilities to concurrently execute heterogeneous (non)MPI  tasks across multiple heterogeneous resources, with tasks implemented as an executable or a Python function. Parsl offers workflow management capabilities, allowing the development of data workflows via its API. RP and Parsl capabilities are delivered with no changes to the application code, allowing users to benefit from Parsl's flexible API and RP scale and performance with no code migration or refactoring efforts.

\begin{figure}[ht]
     \centering
     \includegraphics[width=0.49\textwidth]{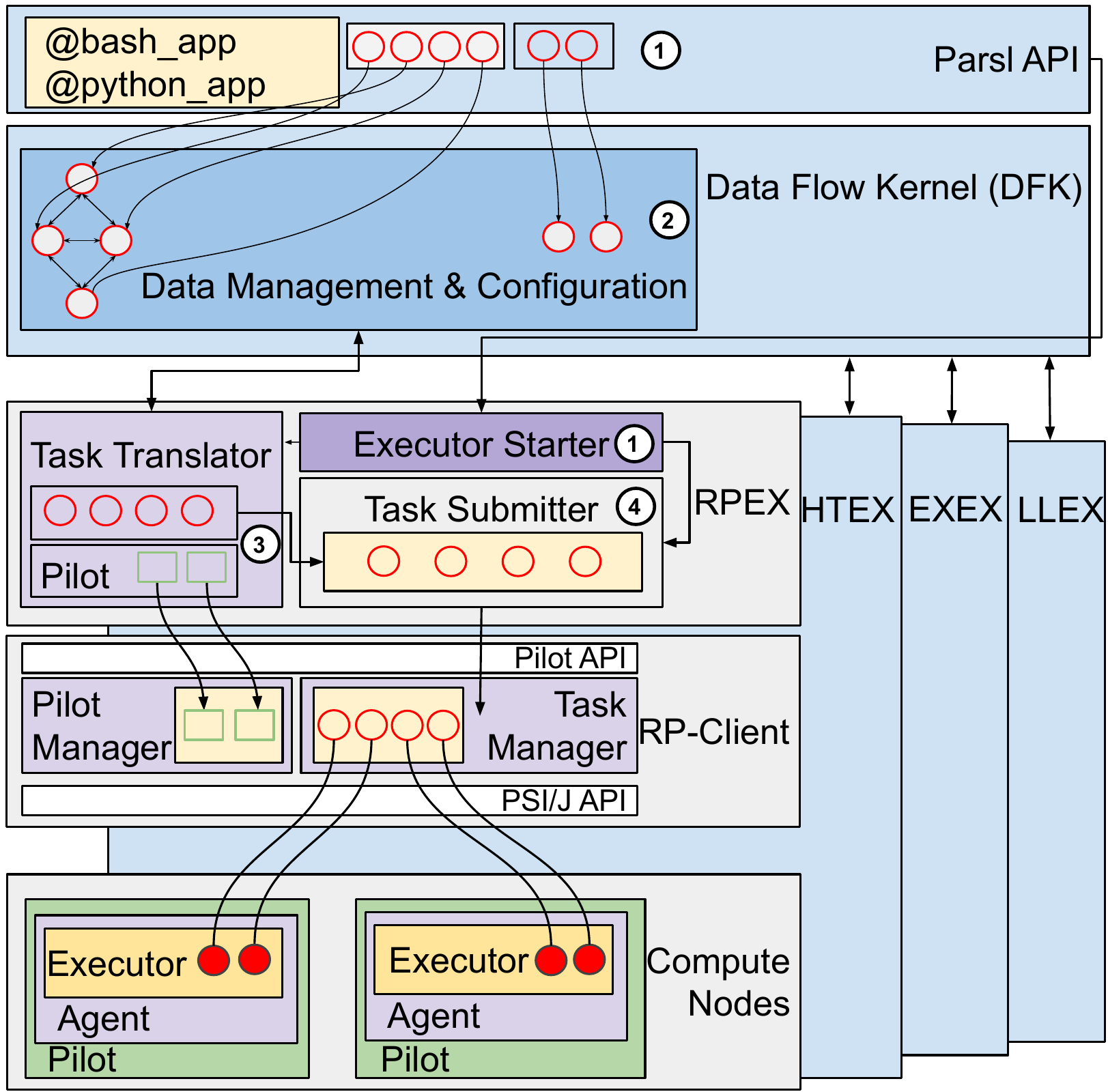}
     \caption{RPEX Architecture. Integration between Parsl (blue boxes) and RADICAL-Pilot (purple and green boxes) via a Task Translator function.}
     \label{fig:rpex}
\end{figure}

RPEX has three main components shown in Fig.~\ref{fig:rpex}: executor starter, task translator, and task submitter. Once Parsl starts the executor~(Fig.~\ref{fig:rpex}~\circled{1}), it submits the resolved tasks directly to RP via the Parsl workflow manager as a ready-to-execute task~(Fig.~\ref{fig:rpex}~\circled{2}). Once RP receives the task, it detects the task type, assigns the resource requirements for each task as specified by the user via~\texttt{parsl\_resource\_specification}~(Fig.~\ref{fig:rpex}~\circled{3}) and submits the task to RP's \texttt{TaskManager}~(Fig.~\ref{fig:rpex}~\circled{4}). Once the tasks are in a state of \texttt{DONE, CANCELED or FAILED}, RP notifies Parsl about the state of the tasks for further processing or for Parsl to declare the execution as done and to shut down the executor.

Integrating RP and Parsl was straightforward except for a difference in resource management. That required aligning RP's task API with Parsl's future tasks by creating a middle point component responsible for translating Parsl's futures into RP's tasks. Most importantly, the task translator's primary duty is to extract the resource requirements from Parsl's tasks and map them to RP's task to enable the use of RP’s resource management capabilities in RPEX.
Beyond coding RP's interface to Parsl's executor API, we wrote unit tests for RPEX and integrated those tests into Parsl's continuous integration infrastructure. Finally, RPEX required documentation by extending both Parsl and RP documentation and adding specific tutorials.

\subsection{Flux Integrations}

We integrated Parsl and RADICAL-Pilot with Flux to add Flux's scheduling and launching capabilities to both systems' launching methods. As seen in \S\ref{sec:components}, Flux is built to enable effective and efficient execution of large-scale executions of tasks. Both Parsl and RADICAL-Pilot can benefit from those capabilities, especially for homogeneous tasks. Concurrent heterogeneous tasks and high-throughput scheduling require executing multiple Flux instances and partitioning the tasks across those instances.

\textbf{Parsl}'s FluxExecutor supports Parsl apps that require complex sets of resources (like MPI or other compute-intensive applications) and collections of applications with highly variable resource requirements. Flux’s sophisticated and hierarchical scheduling makes these applications logically and efficiently executed across Flux-managed resources. Flux is integrated with Parsl via a Python-based wrapper around the Flux API that implements Parsl's Executor interface. We chose to implement the executor interface rather than Parsl's provider interface, which is used for schedulers, as the executor interface provides more fine-grained control over execution. For example, it allows Flux to manage resources dynamically according to the resource specification provided, rather than the simple batch job provisioning offered by the provider interface. Parsl apps are submitted as Flux Jobs to the underlying Flux scheduler via the Executor API. The Flux executor requires a Flux installation to be available locally and located either in PATH or through an argument passed at creation time.

While task execution via the Executor is straightforward, the team had to integrate the Flux resource description model to enable Parsl apps to carry requirements. The integration allows developers to associate a dictionary containing the Flux resource spec with a Parsl app. Supported keys include the number of tasks, cores per task, GPUs per task, and nodes. The Parsl/Flux integration has been used by various users, including for weather modeling workflows at NOAA.

The integration process highlighted several challenges, including diverse resource specifications and issues matching environments. As Parsl supports various executors, each with its way of representing resource specifications (e.g., names of attributes, set of configurable attributes provided), we discussed building a common resource specification. We discussed this approach with various Parsl stakeholders and ultimately decided to support the description used by the executor. This simplifies integration significantly but reduces portability between executors as users must convert resource specifications. We based our decision on the fact that, in several cases, there was no one-to-one mapping between attributes, and we were concerned that users familiar with a particular executor would be confused by using different attributes. However, we identify this as an opportunity for future work.

After completing the integration, we worked with researchers at NOAA to use it for weather simulations. This work aimed to use Parsl to orchestrate a workflow comprised of diverse task types, from single-core processes to MPI tasks. The most significant obstacle was configuring the environments to work cohesively. Using Spack resulted in issues regarding differing Python dependencies between Parsl and Flux. Ultimately, the team used Spack to deploy Flux, used Pip to install Parsl, and then manually installed two Python libraries to specific versions that worked with both Parsl and Flux. Future work in the SDK will focus on establishing compatible environments and regularly testing the pair-wise installations to identify conflicting versions.

\begin{figure}[ht]
  \centering
  \includegraphics[width=0.49\textwidth]{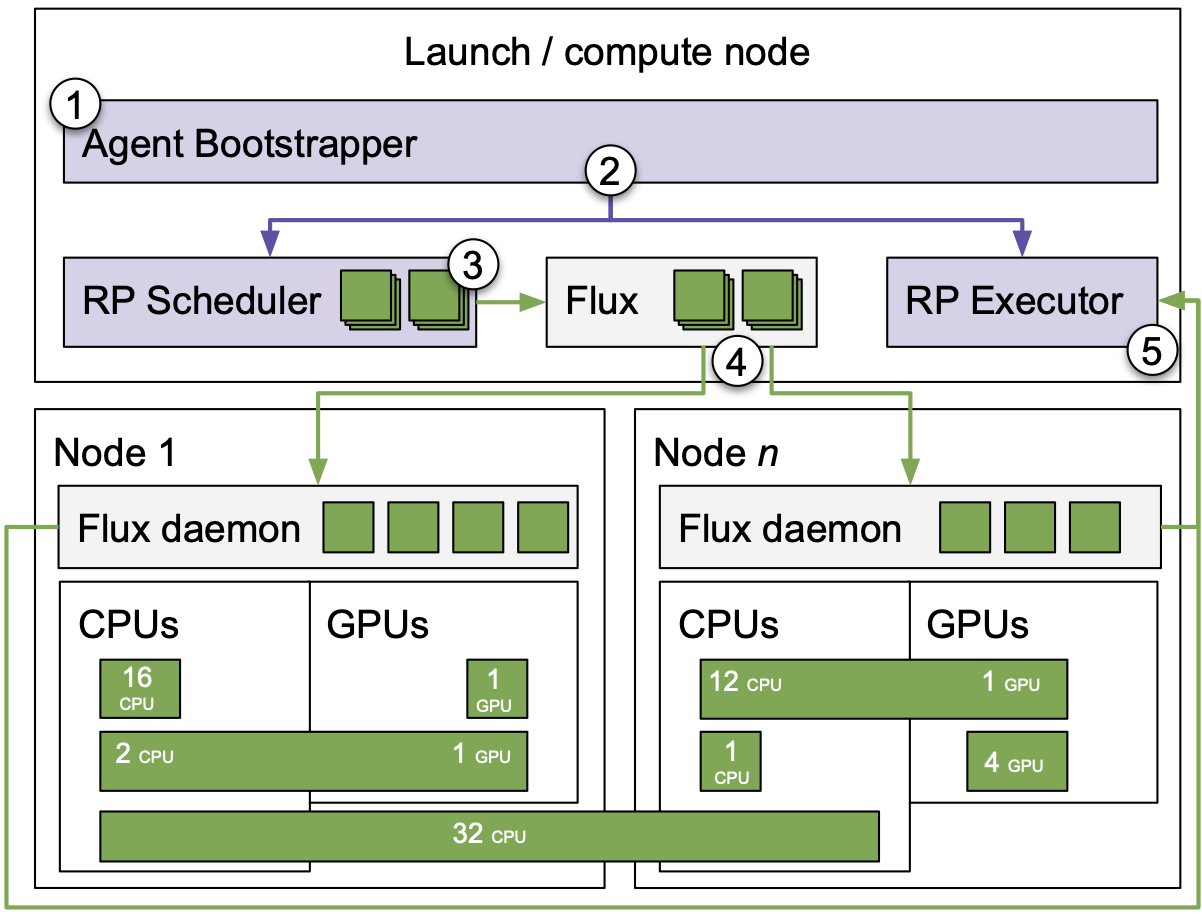}
  \caption{Integration of Flux into RADICAL-Pilot.}
  \label{fig:rp-flux}
\end{figure}

\textbf{RADICAL-Pilot (RP)} and Flux were integrated using the Flux scheduler and task-launching mechanisms as corresponding components within RP. RP provides component-level API, which allows the creation of different types of RP Scheduler, Executor, and Launching Method. Users can use Flux-based components for a corresponding RP application by specifying it in the target platform configuration. RP starts the Flux instance (i.e., Flux-based scheduler and executor) while bootstrapping its components after having obtained the HPC resources allocated via a pilot job (see Fig.~\ref{fig:rp-flux}). Flux schedules, places, and launches tasks on the compute nodes of that allocation via its daemons. RP tracks task completion and makes available this information across its components. If more tasks are available, RP passes them to Flux to execute on the freed resources. Since RP supports multiple instances of components of the same type, RP can increase the overall task throughput by launching multiple Flux instances within the same job allocation and using them concurrently.

Internally, RP launches the Flux instances to execute on a specific subset of compute nodes. The Flux instances are configured by adjusting the Slurm environment settings and, specifically, the node list so that each Flux instance `sees' a different subset of the nodes available in the allocation. Tasks proxied from RP to Flux are then executed on those specific nodes. Task state updates are collected by monitoring the Flux event channel and converting Flux state updates to RP task state updates. Currently, RP implements only basic load balancing between the Flux instances, which can be improved in the future. Flux itself also allows for hierarchical scheduling. In contrast to the approach implemented in RP, Flux hierarchies support only the scheduling of a homogeneous bag of tasks where RP can handle heterogeneous tasks, which is what the workloads that RP typically manages require.

The integration of RP and Flux is a significant step towards enabling the execution of large-scale workflows on HPC platforms. The integration is currently being used to support the execution of exascale workflows on the Frontier supercomputer at Oak Ridge National Laboratory.

\subsection{PSI/J Integrations}

We integrated Parsl, RADICAL-Pilot, and Swift/T with PSI/J, enabling the transparent portability of the applications written utilizing those components across most of the DOE HPC platforms.

\textbf{Parsl} integrates PSI/J by implementing Parsl's extensible \textit{provider} interface in Python (using \texttt{psij-python}). The provider interface requires implementing \texttt{submit}, \texttt{cancel}, and \texttt{status} functions, translating to similar calls in PSI/J. These interfaces are very similar as they are both modern and Python-based; thus, integration was relatively straightforward. The main complexity of the integration stems from the need to translate between Parsl job description objects and those used by PSI/J. However, because the data structures are similar, the implementation is minimal, with roughly 150 lines of code, including logging and error handling. However, the benefits of this integration are significant as it enables the Parsl team to leverage the community-maintained interface to the broad ecosystem of schedulers that are used.

\textbf{RADICAL-Pilot (RP)} uses a \texttt{PilotLauncher} component to interface with the target resource's batch system, to submit pilot jobs, and to monitor their health. That launcher component traditionally uses RADICAL-SAGA~\cite{merzky2015saga} to abstract those interactions. The abstractions provided by PSI/J are conceptually very close to those exposed by SAGA; thus, adding an additional PilotLauncher component based on \texttt{psij-python} required only minor changes to RP code. The PSI/J launcher replaced the SAGA launcher for local submissions to the batch system (see ``PSI/J API'' in Fig.~\ref{fig:rpex}). The main semantic difference to the SAGA backend is the support for remote job submission and, thus, for remote pilot placement in RP. However, the PSI/J API itself, and hence the PSI/J-RP integration, is independent of the backend, and the integration code will seamlessly support remote submissions once PSI/J is extended to support those. The SAGA backend will be retired once PSI/J supports remote submissions.

\textbf{Swift/T} has a single-site model, meaning a Swift/T workflow executes inside a single scheduled job. Longer-running campaigns are typically managed through a checkpoint-restart mechanism (see \S\ref{sec:sucess} for an example with the CANDLE~\cite{wozniak2018candle} use case). Thus, a single scheduler job has to be issued. Constructing this job is a software engineering task that has shared responsibility among three conceptually distinct roles: (1) workflow developer/user, (2) Swift/T maintainer, and (3) HPC site maintainer/administrator. The workflow developer/user simply wants to run scientific workloads using Swift/T and has limited knowledge of Swift/T internals or site-specific technicalities. The Swift/T maintainer understands Swift/T conventions but cannot be responsible for all possible use cases or runtime environments. The HPC site maintainer/administrator installs Swift/T on the site and has in-depth knowledge of the site in question and how to use it. However, it has limited understanding of how Swift/T internals or how its conventions are relevant.

Swift/T was initially packaged with a suite of shell scripts to support running a monolithic Swift/T MPI job using job script templates filtered with settings specified by the user. This model made it easy for advanced users to adjust the scripts as needed quickly. It also posed a code management challenge, as the number of scripts and the variety of computing sites available led to common problems. Swift/T is now bundled with a PSI/J script that fits into this model. Still, all actual interaction with the scheduler is done through PSI/J, pushing the complexity and responsibility of managing scheduler changes and exotic settings into the reusable PSI/J suite. This approach also has the benefit that experienced PSI/J users will likely find Swift/T behavior more understandable.

Swift/T originally used GNU M4~\cite{GNU_M4_WWW} to filter template scripts into submit scripts for the various HPC workload managers. In this approach, an invocation such as \\
\texttt{\$ swift-t -m pbs -n 8 workflow.swift} \\
translates \texttt{workflow.swift} into internal format \texttt{workflow.tic} and issues a PBS job with \\
\texttt{\$ mpiexec -n 8 workflow.tic} \\
The \texttt{-n~8} specifies 8 MPI processes for the job. In practice, there are many other system-related settings that may be provided to \texttt{swift-t} via command-line flags and environment variables. Note that the contents of \texttt{workflow.swift} do not affect the PBS job; the Swift/T architecture separates the internal logic expressed in the workflow language from the system-level specification of the runtime environment. These settings are then translated by simple M4 patterns into the backend script for PBS specified with \texttt{-m~pbs}. In the interest of maintainability, we restrict our M4 usage to only \texttt{m4\_ifelse()} and a custom \texttt{getenv()} macro. Much work goes into maintaining these scripts, as the various settings are expressed differently among the workload managers, and site-specific customization is common at supercomputing facilities.

In the Swift/T-PSI/J integration effort, we enabled the user to simply specify \\
\texttt{\$ swift-t \textbf{-m psij} -n 8 workflow.swift} \\
This invokes a wrapper script \texttt{turbine2psij.py}, bundled with Swift/T, which is a short Python script that does a \texttt{import~psij}, then constructs a \texttt{psij.JobSpec} with the settings. This is then issued to the workload manager with \texttt{psij.JobExecutor.submit()}. Thus, no additional knowledge of workload managers and their specifications is needed on the Swift/T maintainer side. This solves the multi-role software engineering task specified above. The workflow developer/user writes workflows and invokes Swift/T as usual with trivial additional knowledge; any additional knowledge about PSI/J will help with the Swift/T case and other PSI/J uses. The Swift/T maintainer outsources all scripting complexity to the PSI/J maintainers, allowing that role to focus on Swift/T improvements. The HPC site maintainer/administrator, with moderate knowledge of PSI/J and in-depth knowledge of the local site, will be able to easily address any issues with making things work, even without knowledge of Swift/T internals.

\section{Success Stories}\label{sec:sucess}

We propose three exemplar `success stories' of using SDK technologies for different use cases in diverse scientific domains. Those show how SDK enables a wide range of workflow applications and resources while offering consistent packaging, testing, and documentation of those technologies. Note that we present use cases that require individual components of SDK but also their integration. That shows how SDK provided an aggregation venue and added value in the form of new capabilities.

Note that the choice of a specific workflow technology offered by SDK to satisfy a specific use case is socio-technical. That means that each choice is grounded both on a set of specific capabilities that each tool offers but also on more qualitative factors like ongoing collaborations among research groups and PIs, personal preferences, and present or future funding opportunities. As SDK offers tools with overlapping capabilities, often a use case could be supported by more than a single tool. We see that as a positive factor of the SDK approach, which promotes a software ecosystem with complementary technologies that adhere to high engineering standards, are maintained, well documented, and tested on the target infrastructures.

Beyond the three exemplar `success stories' listed below, SDK components and their integrations are being used to support a variety of ongoing use cases. Those use cases span diverse scientific domains and benefit from the documentation, testing, and capabilities offered by SDK. For example, RADICAL-Cybertools is using the integration with Flux to support the execution of exascale workflows on Frontier. Without Flux, RADICAL-Cybertools would be limited in the number of concurrent tasks that could be executed due to the configuration choices made by the Frontier management team. Further, PSI/J is now the default resource management API for Parsl, RADICAL-Cybertools and Swift and, as such, it is used in every use case supported by those systems. Note that, beyond supporting use cases, Exaworks SDK is now also focusing on expanding the number of workflow technologies supported and establish its testing infrastructure as an official capabilities of several DOE labs.

\subsection{Using EnTK for ExaAM workflows}

\begin{figure}[ht]
	\centering
	\includegraphics[width=0.49\textwidth]{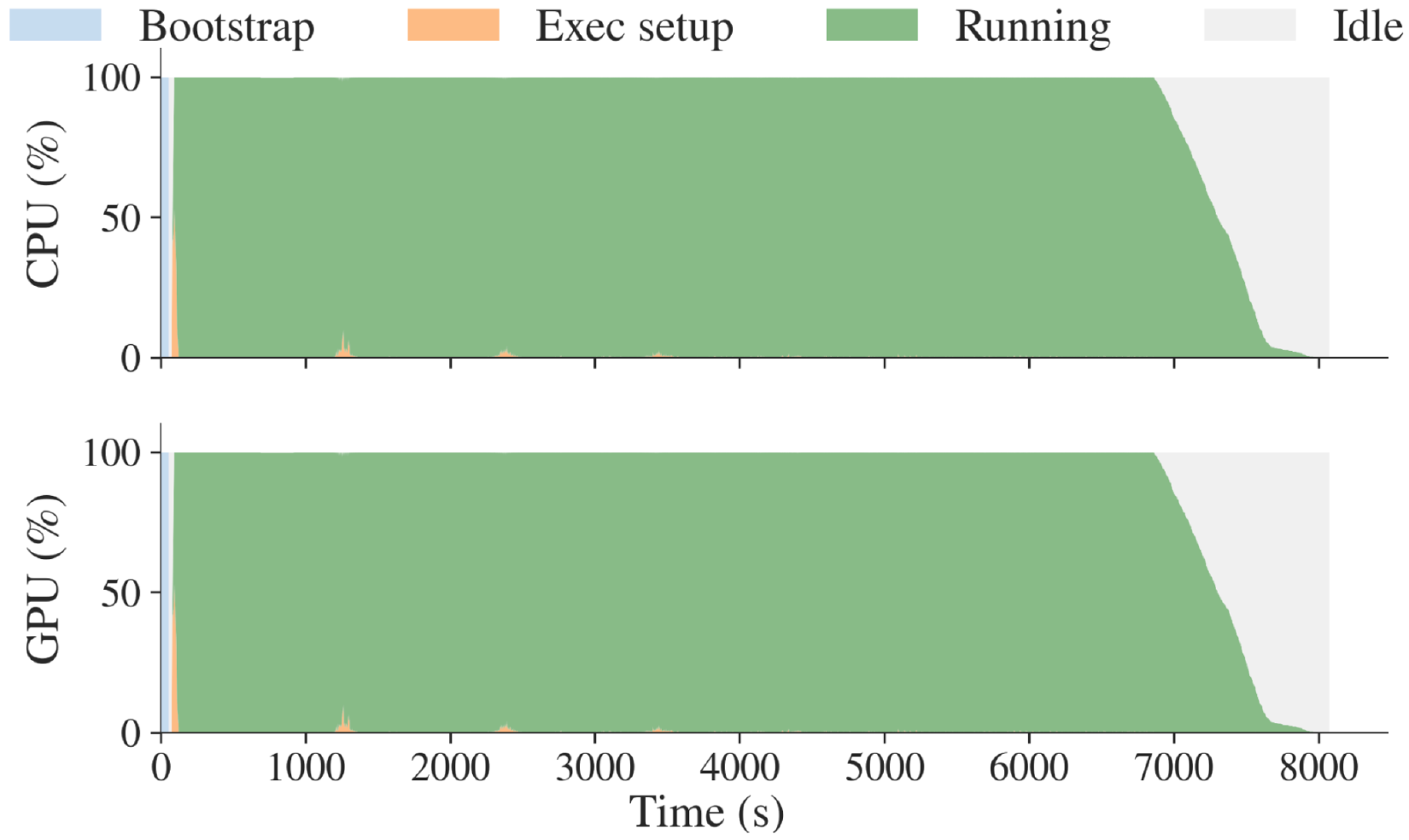}
	\caption{RADICAL Cybertools were used to implement a scalable UQ workflow with the ExaAM team. These plots show overall utilization for the Frontier challenge run was 448,000 CPU cores and 64,000 GPUs, not including 8 CPU cores per node reserved for system processes.}
	\label{fig:exaworks-exaam-utilization}
\end{figure}

The DOE ECP Exascale Additive Manufacturing project (ExaAM) developed a suite of exascale-ready computational tools to model the process-to-structure-to-properties (PSP) relationship for additively manufactured (AM) metal components \cite{exaam-short-paper}. ExaAM built an uncertainty quantification (UQ) pipeline (aka campaign) to quantify uncertainty's effect on local mechanical responses in processing conditions.

ExaWorks teamed up with ExaAM to implement a scalable UQ pipeline solution using ExaWorks SDK components. After initial meetings to elicit the ExaAM workflow requirements, the teams selected the SDK and its RADICAL Cybertools component (see \S\ref{sec:components}) to implement that workflow. The details of the ExaAM workflow are presented in~\cite{exworks-exaam-paper}; below, we offer a summary of the implementation and execution of this workflow at scale.

ExaWorks collaborated closely with the ExaAM team and user(s) to replicate the existing UQ pipeline and its capabilities using SDK's RADICAL Cybertools and, specifically, its workflow engine called EnTK. EnTK enables expressing workflows as pipelines, each composed of a sequence of stages. Each stage contains a set of tasks, enabling concurrent execution, depending on available resources. EnTK's programming model allowed us to directly implement the UQ pipeline into a set of EnTK pipelines without costly and conceptually difficult translations between different workflow representations.

EnTK workflows replaced existing shell scripts that required hands-on management and constant tweaking, made debugging difficult, and consumed resources otherwise available to progress scientific research. SDK and RADICAL Cybertools offered a performant, scalable, automated, fault-tolerant alternative that replicated and enhanced existing capabilities. The developed code is published in the ExaAM project GitHub repository~\cite{exaam_uq_repo}.

We scaled up the EnTK implementation of the ExaAM UQ pipeline on Frontier, utilizing between 40 and 8000 compute nodes for between 2 and 4 hours. Fig.~\ref{fig:exaworks-exaam-utilization} shows a utilization plot for the run on Frontier, demonstrating scalability and utilization. Resource utilization reached 90\% of the 448,000 CPU cores (Fig.~\ref{fig:exaworks-exaam-utilization}(a)) and 64,000 GPUs (Fig.~\ref{fig:exaworks-exaam-utilization}(b)) available for a single run, scaling to the whole Frontier.

\subsection{Enabling CANDLE with Swift/T}

\begin{figure}[ht]
	\centering
	\includegraphics[width=0.49\textwidth]{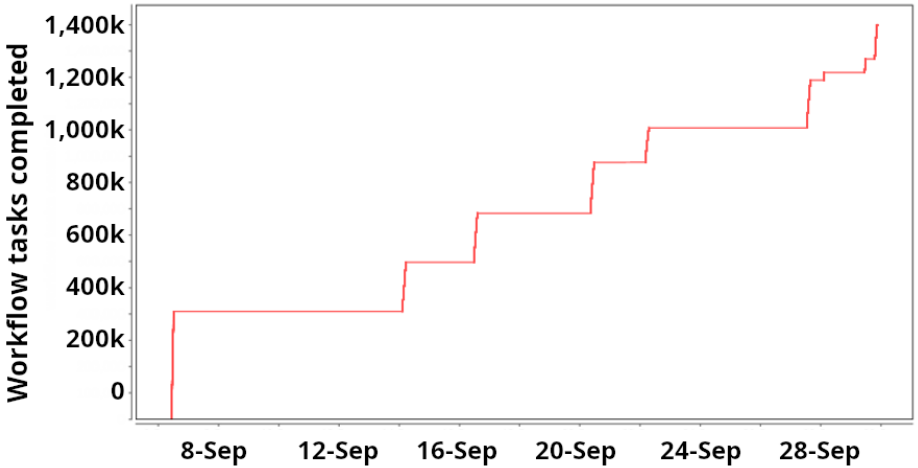}
	\caption{Progress of a typical ``Challenge Problem: Leave One Out'' campaign implemented for the CANDLE team with the Swift/T ExaWorks SDK component.  Several restarts are performed at various scales over a month.  During execution, workflow tasks are rapidly completed. This run was used to look for problems in the training data, and only 2 epochs were run per Uno (see the text) task.}
	\label{fig:swift-t-candle}
\end{figure}

The Swift/T component of the ExaWorks SDK has been developed throughout the project to make it easier to use Swift/T workflows on exascale computers. As part of the SDK effort, we improved Swift/T packaging by developing integration with Docker containers and more maintainable Spack~\cite{gamblin2015spack} and Anaconda packaging scripts. In this integration, Swift/T was used to run a data analysis workflow called ``Challenge Problem: Leave One Out (CPLO)'' developed by ECP/AD/CANDLE on Frontier (see Fig.~\ref{fig:swift-t-candle}). The workflow was previously developed for Summit and ran full scale there, consisting of about 5,000 tasks. The workflow task trains the Uno model, a neural network cancer drug response developed at ANL~\cite{wozniak2020high}. Each task trains the model on a slightly different subset of the Uno data set; common subsets are trained first, and the model weights are reused in fine-tuning tasks by subsequent tasks, resulting in an expanding tree of training tasks. The goal is to study model performance for each record left out at the leaves of the workflow tree (``Leave One Out''). The workflow was expanded for the transition from Summit to Frontier by breaking up the data set further, resulting in an expansion in the number of child tasks per workflow from N=4 on Summit to N=16 on Frontier, resulting in a new workflow tree of size $18*10^6$, a growth factor of 3,600.

Multiple changes were needed to run at the expanded scale on Frontier. Most importantly, the training time for the Uno model was significantly shortened on Frontier, necessitating attention to other parts of the workflow, such as its structure plan file and data subset preprocessing. We applied previously developed MPI-IO-based techniques to stage this data to node-local storage before workflow execution. Checkpointing is part of the workflow at the model and workflow task levels; thus, models can be restarted from the previously saved epoch (a configurable setting), or, if complete, the whole model is reused, and the workflow task is skipped entirely. We typically run at quarter-system-scale on Frontier, with 1 GPU per Uno model, totaling 8 Uno models per node.

\subsection{Colmena with RADICAL-Pilot and Parsl Integration}

Colmena is a Python package for intelligent steering of ensemble simulations~\cite{wardColmena2021}. Colmena is used to steer large-scale heterogeneous MPI Python functions and executable tasks. Further, Colmena uses Parsl to drive an ensemble of computations on HPC platforms to fit interatomic potentials. Often, Colmena applications require executing MPI computations at a modest scale, which, in turn, requires efficiently running many MPI tasks concurrently. Currently, Parsl executors offer limited MPI capabilities, but SDK provides the integration between Parsl and RADICAL-Pilot, one of the RADICAL Cybertools components (see \S\ref{sec:integration}). RPEX, the name given to that integration, satisfies Colmena MPI requirements without requiring changing the existing interface between Parsl and Colmena.

\begin{figure}[t]
	\centering
	\includegraphics[width=0.49\textwidth]{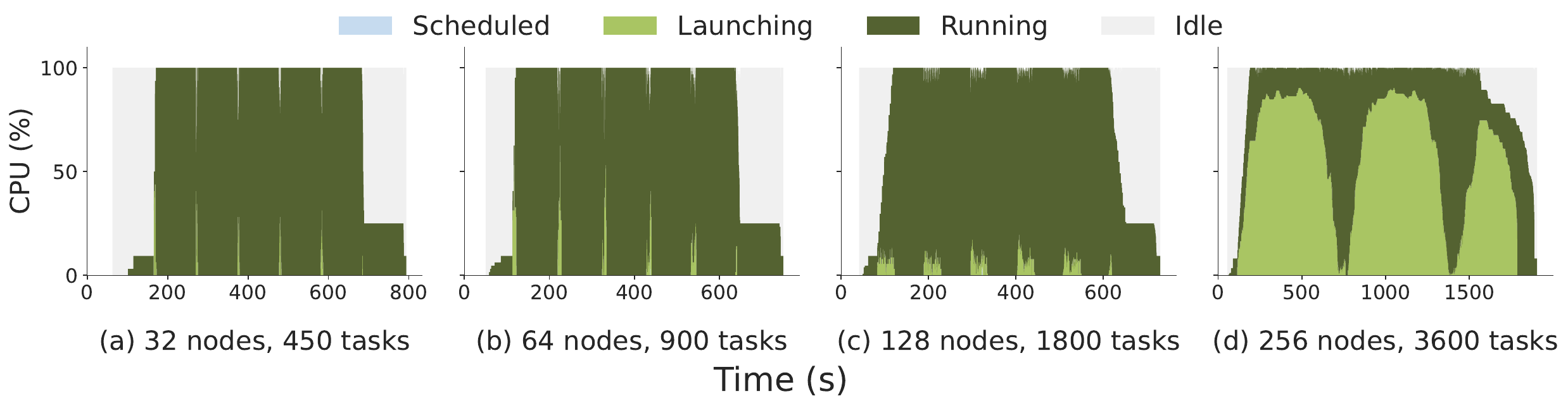}
	\caption{Colmena resource utilization with RPEX (see \S\ref{sec:integration}) on 32, 64, 128, and 256 nodes on TACC Frontera with 56 CPU cores per node.}
	\label{fig:colmena_utilz_rpex}
\end{figure}

RPEX enables Colmena to explore complex multi-physics and multi-scale models by flexibly coupling various types of simulations. This includes seamlessly executing ensembles of tasks with heterogeneous sizes and types, using different computational engines in complex workflows. The provided capabilities are delivered without compromising Colmena's performance and efficiency. We used RPEX in Colmena to execute MPI and Python functions, requiring no changes to the Colmena code base. Fig.~\ref{fig:colmena_utilz_rpex} shows that RPEX reaches a resource utilization of $\sim$99\% while executing both MPI executables and Python functions on up to 256 compute nodes of TACC Frontera, with 56 cores per node~\cite{alsaadiParslRP2022}. While REPEX provides Colmena with new MPI capabilities, RPEX does not introduce additional overheads compared to when Colmena executes only non-MPI functions via Parsl and without RP. In Ref.~\cite{wardColmena2021}, Fig. 3 shows resource utilization comparable to the one achieved with RPEX and showed in Fig.~\ref{fig:colmena_utilz_rpex}.

\section{Testing}\label{sec:testing}

One of the most challenging problems of testing software on HPC platforms is their heterogeneity: tests on one HPC cluster are not necessarily reproducible on other HPC clusters. This difficulty stems from several reasons, including hardware differences, the use of heavily optimized cluster-specific library implementations, and differences in software and configuration. Specific projects, such as PSI/J~\cite{psij-23}, reduce this difficulty by providing a uniform API for accessing parts of the software/library stack. Still, limitations exist to how much one can abstract without compromising performance and required specificity. Thus, a natural solution is to test software on an exhaustive set of HPC platforms. However, such a strategy is generally met with different difficulties since HPC clusters are often found under separate administrative domains with little to no infrastructure that would allow a software package development team to reach a reasonable subset of them. Typically, the result is that the average HPC software package is tested on a small set of platforms that the development team can muster access to, with most testing on other platforms being delegated to users in an ad-hoc fashion. It is important to distinguish here between unit and integration testing. We assume (and encourage, as part of participation in the SDK) that each SDK component contains appropriate unit-level tests.

\subsection{Testing Infrastructure Overview}

The ExaWorks SDK addresses the testing problem by providing an infrastructure that enables testing of the SDK components on DOE platforms and other platforms that users can access. The infrastructure consists of a test runner framework and a dashboard.

The test runner framework is a collection of tools and practices that enable the deployment of SDK components and the execution of tests therein. The main drivers for the framework are GitLab Continuous Integration pipelines deployed at various DOE labs. In addition, GitHub actions and direct invocation (which can be driven by the \texttt{cron} tool) are also supported and used. Each pipeline consists of a deployment step and a test execution step. Three means of deployment are provided: \texttt{pip}, \texttt{conda}, and \texttt{spack}, but the support varies by the package being tested and the location where the pipeline is run. A location-agnostic version of the SDK is also provided as a \texttt{Docker} container. Active pipelines are configured for ALCF/ANL (Polaris), LLNL (Lassen, Quartz, and Ruby), NERSC (Perlmutter), and OLCF/ORNL (Ascent and Summit). The test execution step invokes a set of validation tests for each SDK component and integration tests that validate the correct interfacing of two or more SDK components when appropriate.

A testing infrastructure would not be complete without the means to collect and meaningfully present results to developers and (potential) users. The ExaWorks SDK testing dashboard addresses this piece of the testing puzzle: the mechanism to report test results to developers and, in its final production version, also to users. The testing dashboard was initially developed for the PSI/J-Python project, where it enabled the centralization of results of user-maintained test runs. Like PSI/J-Python, the ExaWorks SDK adopts a hybrid approach, with tests on some HPC systems maintained by the ExaWorks team while allowing users to set up test runs on machines under their control.

\begin{figure}[t]
	\centering
	\includegraphics[width=0.49\textwidth]{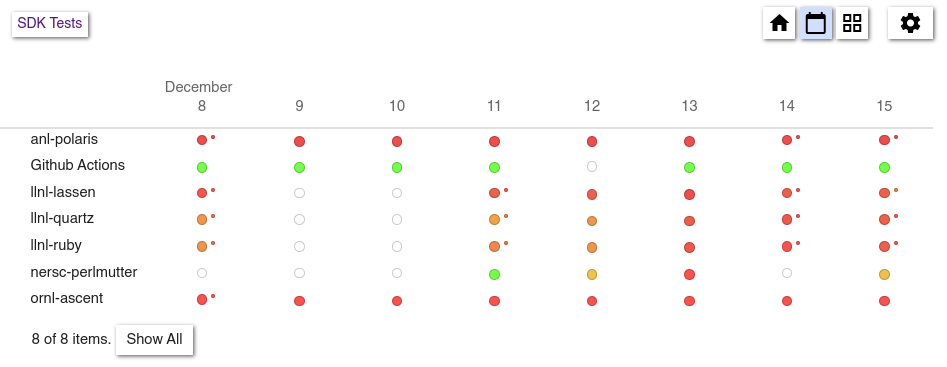}
	\caption{The SDK Testing Dashboard Summary Page}
	\label{fig:dashboard-summary}
\end{figure}

The testing process is as follows. A client-side test runner (typically invoked by a GitLab pipeline) executes desired tests and captures the result and ancillary information, such as output streams or logs. This information is then uploaded to the testing dashboard to present the results to users and developers. By default, results are aggregated by site and date into a calendar view---see Figure~\ref{fig:dashboard-summary}, which shows a quick overview of the overall pass/fail trends over the last few days. Users then can select sites and navigate to specific test runs, which allows them to examine individual test results and client-provided outputs and logs.

The testing dashboard consists of a backend and a frontend. The backend provides authentication, stores test results in a database, and responds to queries for historical test data; the frontend is a web application implemented using the Vue.js~\cite{vuejs-web} library and displays the test information to developers and users alike. Like the PSI/J-Python testing dashboard, the SDK dashboard uses a simple authentication mechanism that requires a verified email address. This is done to associate result uploads with an identity that can be used to manage access to the dashboard. The choice to use simple email validation rather than an authentication provider service is motivated by a desire to allow test results from users from any institution and private users.

\subsection{Challenges}

Our approach is not without challenges, some of which are essential to the problem we are attempting to address, while others are consequences of choices we made along the way. One of such difficulties, which results from essential and circumstantial complexities, is that SDK package tests are done with all SDK packages installed. That is because many of the SDK packages are meant to be interoperable, as well as the assumption that in many realistic environments, it is entirely possible or even desirable to have multiple arbitrary SDK packages installed on one system. This can lead to dependency conflicts that individual package maintainers cannot reasonably or effectively address.

A related but distinct difficulty is installing diverse packages on heterogeneous computing platforms. In many cases of actual and possible component packages of the SDK, compilation on target compute resources is necessary, with exceptions for Python projects with no native (compiled) parts and other projects written exclusively in interpreted languages. The ExaWorks SDK supports two solutions for compiling and installing packages: Conda~\cite{conda-web} and Spack~\cite{gamblin+:sc15}. Member projects of the SDK must either provide a Conda package specification or be available through Spack (or both). Many projects, however, lack the resources to maintain Conda or Spack packages, and providing such packages as a condition for inclusion in the ExaWorks SDK represents an additional cost. Furthermore, both Conda and Spack are designed to create a sandbox in which specific dependencies are compiled and installed, such as to not conflict with system packages, some of which are optimized specifically for the target system. Although both Conda and Spack allow exceptions to be made and system packages to be used directly, a trade-off is forced: run the risk of incompatibilities by using an optimized dependency provided by the system or sacrifice performance both at run-time as well as during installation by compiling and installing unoptimized dependencies.

Finally, testing on multiple HPC platforms managed by diverse organizations poses sociotechnical issues related to supporting those tests on those resources, resource allocation and scheduling, and security policies. Often, in the DOE space, tests need to be audited and approved before being routinely executed on HPC platforms. Further, routinely running tests requires dedicated resources associated with a project and a related allocation. Without specific policies and agreements about resources and allocations dedicated to testing within each institution, ExaWorks runs the SDK tests utilizing the allocation of the project. While that worked for the project's duration, it does not allow the establishment of a long-term testing strategy for SDK within the DOE community.

During our work on a testing infrastructure, we encountered several practical obstacles. Our first (chronologically) obstacle was access to HPC systems, especially when a proposed ECP project-wide testbed failed to materialize fully. Our team was left with submitting individual project applications for each HPC center where tests were to be deployed. While we could apply for startup allocations without much effort, none of the centers had provisions for allocations that were meant to address infrastructure testing. Much effort was spent deciphering the required GitLab configurations for each HPC center, with specific and complex configuration files needed for each machine. This could be seen as a success story since individual software package maintainers would have to go through a similar process to run tests on these systems, and the ExaWorks SDK has the potential to tame this difficulty.

\section{Tutorials and Documentation}\label{sec:docs}

Alongside testing, documentation is also a fundamental element of the ExaWorks  SDK. SDK documentation has to satisfy three main requirements: (1) centralize into a single venue and under a consistent interface all the information specific to SDK; (2) avoid duplication of documentation between SDK and its tools; (3) minimize maintenance overheads; (4) manage a rapid rate of obsolescence in the presence of continuously and independently updated software components; (5) use the same documentation for multiple purposes like training, dissemination, hackathons, and tutorials.

Documenting the SDK poses specific challenges compared to documenting a single software system. The SDK is a collection of software components independently developed by unrelated development teams. While each tool is part of SDK, the ExaWorks team does not participate in the development activities supporting each component, decide when and how those components are released, and how each component's documentation is updated or extended. That has the potential to make the SDK documentation obsolete, very resource-intense to maintain, and prone to create duplication detrimental to the end user.

We consistently scoped the SDK documentation to offer static and dynamic information. Static information focuses on the SDK itself, offering the needed information about what it is and it is not, the list of current components, the minimum requisites for a component to be part of the SDK, the process to follow to include that component, and details about code of conduct and governance. Due to the challenges listed above, we avoid static information about the SDK components, directly linking specific documentation for each tool. As such, we provide a hub where users can find a variety of pointers to the vast and often dispersed software ecosystem to support the execution of scientific workflows at scale on DOE HPC platforms.

Notwithstanding the availability of tool-specific documentation, we had to document the use of those tools on a specific set of DOE platforms and, especially, as a set of integrated systems. Thus, we devised a novel approach to dynamic documentation centered on tutorials designed to be used for outreach, training, and hackathon events. At the core of our approach are Jupyter notebooks containing both documentation and code to deliver: (1) paradigmatic examples of scientific applications developed with the SDK components and executed on DOE platforms; (2) tutorials about capabilities of SDK that serve the specific requirements of the DOE workflow community; (3)  detailed examples of resource acquisition, management, tasks definition and scheduling; (4) debugging and tracing workflow executions; and (5) many other common tasks required by executing workflows on HPC platforms.

The SDK tutorials avoid overlapping with the tutorials specific to each SDK component, focusing on documenting the use of SDK within the DOE software ecosystem and the integration among SDK components. Such integrations are not specific to any tool, so SDK documentation represents an ideal venue for collecting and organizing that information. Further, SDK tutorials are maintained and distributed via GitHub to make them available to other projects supporting the development of DOE platform workflow applications. GitHub enables contributions from the whole community, creating a single body of encoded information that can be maintained and updated beyond the end of the ExaWorks project.

Organizing the dynamic component of SDK documentation on GitHub allowed us to extend its use beyond its traditional boundaries. Utilizing GitHub workflows, we created a continuous integration platform where each Jupyter notebook can be automatically executed every time any SDK component is released. That avoids manually maintaining all the tutorials and makes it immediately apparent when a new component release breaks the tutorial's code. Further, the same tutorials can be seamlessly integrated with Readthedocs~\cite{sdk-readthedocs}, the system we use to compile and distribute the SDK documentation. Executing the tutorials every time the documentation is published guarantees that the tutorials work, greatly improving the quality of the information distributed to the end user. Finally, as part of the GitHub workflows, we package all the tutorials into a Docker container~\cite{sdk-container} that enables users to execute the tutorials both locally or via Binder~\cite{binder-web} with minimal overheads and portability issues.

\section{Conclusions}\label{sec:conclusions}

Our experience developing the ExaWorks SDK offers valuable insight. There is a gap in the DOE software ecosystem to support the execution of scientific workflows at scale. On the one hand, many middleware components are required to execute those workflows; on the other hand, DOE maintains diverse HPC platforms with different capabilities, policies, and support levels.

Users face the challenge of selecting a set of middleware components to obtain an end-to-end software stack that works on one or more of those resources. That choice is challenging because middleware components have overlapping capabilities, work only on a subset of platforms, may work with some other components, or require the user to lock into a specific software stack. Further, users do not know whether and how well each component is tested, internally (unit tests) and on a specific DOE platform (integration tests). Finally, users must patch together component-specific documentation without tutorials designed to illustrate how to execute workflows at scale on a given DOE platform. Ultimately, users face a steep learning curve and time-consuming testing and debugging of diverse middleware components.

As seen in \S\ref{sec:components}, ExaWorks SDK fills that gap by providing a curated set of components that: (1) span the software stack required to execute scientific workflows on DOE HPC platforms; (2) comply with software engineering best practices, including testing coverage, continuous integration, well-defined APIs, and comprehensive documentation; and (3) are maintained, open source and widely adopted. These properties are proven to satisfy the requirements elicited from the DOE workflow community and represent a valuable guideline for future SDKs.


Reducing the cost and effort duplication that has produced a slew of comparable software tools with overlapping capabilities requires enabling their integration, independent of the development team that develops and maintains each component. As shown in \S\ref{sec:integration}, ExaWorks SDK has proven that such an integrative approach works, requires minimal development effort, and provides added value. \S\ref{sec:sucess} detailed how SDK components delivered necessary capabilities to diverse scientific domains, both stand-alone and integrated.

Beyond curating and integrating a set of suitable components, the ExaWorks SDK reaffirms the importance of both testing (\S\ref{sec:testing}) and documentation (\S\ref{sec:docs}). ExaWorks delivered novel approaches and technologies in both domains, showing that continuous integration on the HPC platforms has become necessary.  With the growing importance of workflows as a scientific application paradigm, users must be given tangible proof that a middleware stack will not fail them when deployed on a specific HPC platform. Gone are the days when users had the time and resources to work as beta testers or could stand months-long delays before executing a computational campaign. Similarly, documentation is not a useful but an ancillary add-on to the software tools. Building deep stacks of middleware components is becoming increasingly complex, and, again, users do not have the time or resources to `try it out' and explore the problem space by trial and error.

Overall, our experience with SDK indicates that DOE must commit resources to create, maintain, and integrate a curated set of workflow technologies in the long run. The DOE computing facilities must officially support those technologies, and those need to go beyond just the HPC platforms. Continuous integration platforms and git-based repositories integrated with workflow capabilities, alongside virtualization platforms, are becoming necessary components of an HPC platform and require careful design and analysis from a performance point of view, as with any other system of an HPC facility. ExaWorks SDK has paved the way, showing the importance of those capabilities and offering tools, policies, and prototypes that can be readily used and extended. In that direction, follow-up projects like the Partnering for Scientific Software Ecosystem Stewardship Opportunities (PESO)~\cite{peso-web} or the Center for Sustaining Workflows and Application Services (SWAS)~\cite{swas-web} are the natural successors of ExaWorks and are already leveraging the work done with the ExaWorks SDK.

\section*{Acknowledgements}

This research was supported by the Exascale Computing Project (17-SC-20-SC), a collaborative effort of the U.S. Department of Energy Office of Science and the National Nuclear Security Administration. This work was performed under the auspices of the U.S. Department of Energy by Lawrence Livermore National Laboratory under Contract DE-AC52-07NA27344 (LLNL-CONF-826133), Argonne National Laboratory under Contract DE-AC02-06CH11357, and Brookhaven National Laboratory under Contract DESC0012704. This research used resources of the OLCF at ORNL, which is supported by the Office of Science of the U.S. DOE under Contract No. DE-AC05-00OR22725. OSPREY's work was supported by the National Science Foundation under Grant No. 2200234, the National Institutes of Health under grant R01DA055502, and the DOE Office of Science through the Bio-preparedness Research Virtual Environment (BRaVE) initiative.
This manuscript has been authored in part by UT-Battelle, LLC, under contract DE-AC05-00OR22725 with the US Department of Energy (DOE). The publisher, by accepting the article for publication, acknowledges that the U.S. Government retains a non-exclusive, paid-up, irrevocable, worldwide license to publish or reproduce the published form of the manuscript or allow others to do so for U.S. Government purposes. The DOE will provide public access to these results following the DOE Public Access Plan (http://energy.gov/downloads/doe-public-access-plan). All the authors contributed to discussing and reviewing the structure and content of the paper. Hategan-Marandiuc and Titov wrote \S\ref{sec:testing}; Alsaadi, Chard, Merzky, Titov, Turilli, and Wozniak contributed to \S\ref{sec:components} and \S\ref{sec:integration}. Turilli wrote the rest of the paper. All the other authors contributed to the software activities described in the paper. Turilli and Jha co-organized the paper.

\Urlmuskip=0mu plus 1mu\relax
\bibliographystyle{IEEEtran}
\bibliography{references}


\end{document}